\documentclass[aps,prc,twocolumn,amsfonts,amsmath,amssymb,showpacs,superscriptaddress,floatfix]{revtex4-1}

\usepackage{graphicx}
\usepackage{longtable}
\usepackage{bm}
\usepackage{multirow}
\usepackage{hyperref}
\usepackage{ulem}
\hypersetup{colorlinks,citecolor=blue,filecolor=blue,linkcolor=blue,urlcolor=blue}
\usepackage{color}
\usepackage{dcolumn}

\begin{document}

\title{High-{\boldmath$j$} neutron excitations outside {\boldmath$^{136}$}Xe}

\author{R.~Talwar}
\affiliation{Physics Division, Argonne National Laboratory, Argonne, Illinois 60439, USA}
\author{B.~P.~Kay}
\email[E-mail: ]{kay@anl.gov}
\affiliation{Physics Division, Argonne National Laboratory, Argonne, Illinois 60439, USA}
\author{A.~J.~Mitchell}
\affiliation{Department of Nuclear Physics, Research School of Physics and Engineering, The Australian National University, Canberra, ACT 2601, Australia}
\author{S.~Adachi}
\affiliation{Research Center for Nuclear Physics (RCNP), Osaka University, Ibaraki, Osaka 567-0047, Japan}
\author{J.~P.~Entwisle}
\affiliation{School of Physics and Astronomy, University of Manchester, Manchester M13 9PL, United Kingdom}
\author{Y.~Fujita}
\affiliation{Research Center for Nuclear Physics (RCNP), Osaka University, Ibaraki, Osaka 567-0047, Japan}
\author{G.~Gey}
\affiliation{Research Center for Nuclear Physics (RCNP), Osaka University, Ibaraki, Osaka 567-0047, Japan}
\author{S.~Noji} 
\affiliation{Research Center for Nuclear Physics (RCNP), Osaka University, Ibaraki, Osaka 567-0047, Japan}
\author{H.~J.~Ong}
\affiliation{Research Center for Nuclear Physics (RCNP), Osaka University, Ibaraki, Osaka 567-0047, Japan}
\author{J.~P.~Schiffer}
\affiliation{Physics Division, Argonne National Laboratory, Argonne, Illinois 60439, USA}
\author{A.~Tamii}
\affiliation{Research Center for Nuclear Physics (RCNP), Osaka University, Ibaraki, Osaka 567-0047, Japan}
\date{\today}

\begin{abstract}

The $\nu0h_{9/2}$ and $\nu0i_{13/2}$ strength at $^{137}$Xe, a single neutron outside the $N=82$ shell closure, has been determined using the $^{136}$Xe($\alpha$,$^3$He)$^{137}$Xe reaction carried out at 100~MeV. We confirm the recent observation of the second 13/2$^+$ state and reassess previous data on the 9/2$^-$ states, obtaining spectroscopic factors. These new data provide additional constraints on predictions of the same single-neutron excitations at $^{133}$Sn.

\end{abstract}


\maketitle

\section{Introduction}

In this work we report on the $j^{\pi}=9/2^-$ and \mbox{$j^{\pi}=13/2^+$} excitations outside of $^{136}$Xe, and discuss them in context of the evolution of these single-neutron excitations across the $N=83$ isotonic chain. Together with previous data~\cite{Kay08}, this provides a consistent description of these orbitals from $^{137}$Xe ($Z=54$) to $^{145}$Sm ($Z=62$).

The character of single-neutron excitations outside of $N=82$ has been explored with nucleon transfer reactions, both in terms of the energy centroid of their strength and the fragmentation of this strength among the actual states of the nucleus.

Booth, Wilson, and Ipson~\cite{Booth74} observed two $\ell=6$, 13/2$^+$ states populated by the ($d$,$p$) reaction on $^{138}$Ba, $^{140}$Ce, $^{142}$Nd, and $^{144}$Sm. These measurements, among others, also identified two $\ell=5$, 9/2$^-$ states outside each of the stable even $N=82$ isotones. Heyde {\it et al.}~\cite{Heyde75} and others~\cite{Isacker79,Trache93,Oros95} explored this fragmentation in terms of coupling to core vibrational states. The spectroscopic overlaps of these same high-$j$ excitations were reassessed using the ($\alpha$,$^3$He) reaction at 51~MeV~\cite{Kay08}. The ($\alpha$,$^3$He) reaction is better matched for higher angular momentum transfer and yields more reliable spectroscopic factors~\cite{Schiffer12,Schiffer13}. These studies of the location of the $\nu0h_{9/2}$ and $\nu0i_{13/2}$ centroids were prompted by new calculations exploring the evolution of single-particle excitations due to  the tensor interaction~\cite{Otsuka05}.

Extending the systematic study of the $N=83$ isotones with transfer reactions to $^{137}$Xe has proven challenging due to xenon being a gas at room temperature. Early measurements using gas targets~\cite{Schneid66,Moore68} did not have the necessary beam energy, nor the most suitable choice of reaction, to successfully probe the high-$j$ 9/2$^-$ and 13/2$^+$ states. The measurement~\cite{Kraus91} of the ($d$,$p$) reaction on $^{136}$Xe in inverse kinematics, the first of its kind, focused on the low-$j$ states and again, had insufficient incident energy to probe the high-$j$ states quantitatively. 

The work of Allmond {\it et al.}~\cite{Allmond12} explored sub-barrier heavy-ion transfer reactions on both $^{136}$Xe and $^{134}$Te. Using the particle-$\gamma$ coincidence technique, they observed the 13/2$^+_1$ state in $^{137}$Xe for the first time. This guided the analysis of a subsequent measurement of the $^{136}$Xe($d$,$p$)$^{137}$Xe reaction carried out in inverse kinematics at 10~MeV/u~\cite{Kay11}. Spectroscopic factors were determined for the two 9/2$^-$ states that share the $\nu0h_{9/2}$ strength and also the lowest 13/2$^+$ state. However, the excitation energy and strength of the  higher-lying 13/2$^+$ state was not observed.

Most recently, Reviol {\it et al.}~\cite{Reviol16} observed the 13/2$^+_2$ state in $^{137}$Xe for the first time. This was done via neutron transfer induced by $^9$Be and $^{12}$C at energies below the Coulomb barrier using the particle-$\gamma$ coincidence technique. Using these data in conjunction with shell-model calculations, they proposed a notably lower value of the $\nu0i_{13/2}$ single-neutron energy at $^{133}$Sn of 2366~keV. Contrary to previous predictions, this value lies {\it below} the neutron separation energy. Their estimate was largely based on the calculations reproducing the energies of the two, now observed, 13/2$^+$ states in $^{137}$Xe and other structurally relevant states in $^{134}$Sb and $^{135}$Sb. 
 
A quantitative determination of the spectroscopic factors for the neutron 9/2$^-$ and 13/2$^+$ excitations in $^{137}$Xe is still lacking. In the present paper, we report on a study of the $^{136}$Xe($\alpha$,$^3$He)$^{137}$Xe reaction at a beam energy of 100~MeV to probe the $\ell=5$, 9/2$^-$ and $\ell=6$, 13/2$^+$ single-neutron excitations. An additional measurement of the $^{144}$Sm($\alpha$,$^3$He)$^{145}$Sm reaction was included in this study to provide a consistency check between this work and that done at a lower incident $\alpha$-particle energy of 51~MeV~\cite{Kay08}. The present data allow for better predictions of the location of the $\nu0h_{9/2}$ and $\nu0i_{13/2}$ strength at $^{135}$Te and $^{133}$Sn, and a measure of the evolution of these single-particle excitations and the influence of the tensor interaction on the neutron single-particle states as the proton orbits are filling.

\section{The experiment}
\label{sec2}

We carried out this measurement, experiment E453, at the Research Center for Nuclear Physics at Osaka University, Japan. An $\alpha$-particle beam of 100~MeV was delivered, via the so-called WS course~\cite{Fujita97,Wakasa02} to the scattering chamber of the Grand Raiden (GR) spectrometer~\cite{Fujiwara99}. The dispersion-matching capabilities of that beam line were not utilized during this experiment. The outgoing ions from the reaction were momentum analyzed by the GR spectrometer and their position and identity determined with vertical drift chambers and scintillators at the focal plane. The ($\alpha$,$^3$He) reaction was carried out on three targets. The first, a 93.8\%-enriched $^{144}$Sm target, prepared from its oxide, with a nominal thickness of 525~$\mu$g/cm$^2$ mounted on a carbon backing of $\sim$40-$\mu$g/cm$^2$ thickness. As mentioned above, this target was used to provide a comparison with a previous measurement of the $^{144}$Sm($\alpha$,$^3$He)$^{145}$Sm reaction carried out at 51~MeV~\cite{Kay08}. 

The second was the $^{136}$Xe target, enriched to greater than 99.9\%, prepared in a manner identical to that described in Ref.~\cite{Entwisle16}, using the RCNP gas-cell target system~\cite{Matsubara12}. The gas-cell had PEN windows~\cite{PEN}, a compound of only carbon, oxygen, and hydrogen isotopes. Such a choice, over other common plastics that often contain nitrogen or chlorine isotopes, was essential for this measurement in terms of leaving the measured spectra relatively free of contaminant peaks. Such windows have been used previously in charge-exchange and transfer-reactions studies on Xe isotopes~\cite{Puppe11,Entwisle16}. The windows of the gas cell were 4-$\mu$m thick. Several cells were prepared in case of failure. Typical $\alpha$-particle beam currents were around 40~enA and the beam had an approximate diameter of $\lesssim$2~mm. Over the course of the measurement, two gas cells were used. The first cell survived approximately $9\times10^{15}$ $\alpha$ particles at the nominal current, and the second receiving about $5\times10^{15}$ (to the end of the experiment). The pressure and temperature of the gas volume were recorded throughout the measurement. Both remained within a few percent of their initial values, being around 16--17~kPa and 298~K, with and without beam. The target had an effective $^{136}$Xe-gas thickness of $\sim$850(120)~$\mu$g/cm$^2$, which was derived independently from both elastically scattered $\alpha$-particle yields, described below, and from an empirical formula given in Ref.~\cite{Matsubara12}. The two results were consistent at the 30\% level, which in turn reflects the uncertainties on the measured absolute cross sections.

Finally, a natural carbon target, of thickness $\sim$10~mg/cm$^2$, was used at each of the angles to observe the contaminant peaks, of which, those from reactions on carbon were dominant. Residual oxygen in the carbon foils gave some indication of the location of the oxygen contaminant peaks. For the measurements on the carbon targets, the optical elements of the GR spectrometer remained the same as those for the respective Sm and Xe target.

The reaction was carried out at angles of $\theta_{\rm lab}=4^{\circ}$, 6$^{\circ}$, 9$^{\circ}$, 12$^{\circ}$, 15$^{\circ}$, 18$^{\circ}$, and 21$^{\circ}$ on the Xe and C targets. Only angles of 6$^{\circ}$ and 9$^{\circ}$ were used for the $^{144}$Sm target. The aperture of the GR spectrometer was 1.36~msr, subtending approximately 1.5$^{\circ}$ horizontally in the laboratory frame. Faraday cups were used to measure the beam current. For all spectrometer angles greater than $\theta_{\rm lab}=4^{\circ}$, a Faraday cup in the scattering chamber (SC cup) was used. At $\theta_{\rm lab}=4^{\circ}$ the SC cup obscured the GR spectrometer aperture. Hence at this angle, a cup placed just after the first quadrupole element of the spectrometer (Q1 cup) was used. This introduced a small systematic uncertainty which was estimated to be less than a few percent.

While absolute cross sections are not required to determine the centroids of single-particle strength, they can be of value in the analysis process by providing comparisons between different targets. Typically, absolute cross sections can be acquired by calibrating the target thickness and spectrometer aperture to well-known reaction cross sections, such as elastic scattering in the Rutherford regime. Due to the thick gas-cell target, such measurements were not possible here. 

Absolute cross sections were acquired by normalizing the yield from ($\alpha$,$\alpha$) elastic scattering at an angle of $\theta_{\rm lab}=6^{\circ}$ ($\theta_{\rm lab}=9^{\circ}$ for the $^{144}$Sm target) to cross sections from optical-model calculations. At such forward angles the cross section is very senstive to the scattering angle and is estimated to be 62(12)\% [33(14)\% for $^{144}$Sm] of the Rutherford scattering cross section, where the uncertainties take into account the variations between different parameters, which are relatively small, and uncertainties relating to those of the angle and aperture, which dominate. 

The elastic scattering on $^{136}$Xe was measured at all of the angles for which the ($\alpha$,$^3$He) reaction was measured, with the exception of $\theta_{\rm lab}=4^{\circ}$. The general features of the scattering data were well reproduced by the optical-model calculations, although the uncertainties in the absolute cross sections are relatively large, being around 30\% for Xe and around 50\% for Sm. 

\section{Analysis and Results}

\begin{figure}[h]
\centering
\includegraphics[scale=0.85]{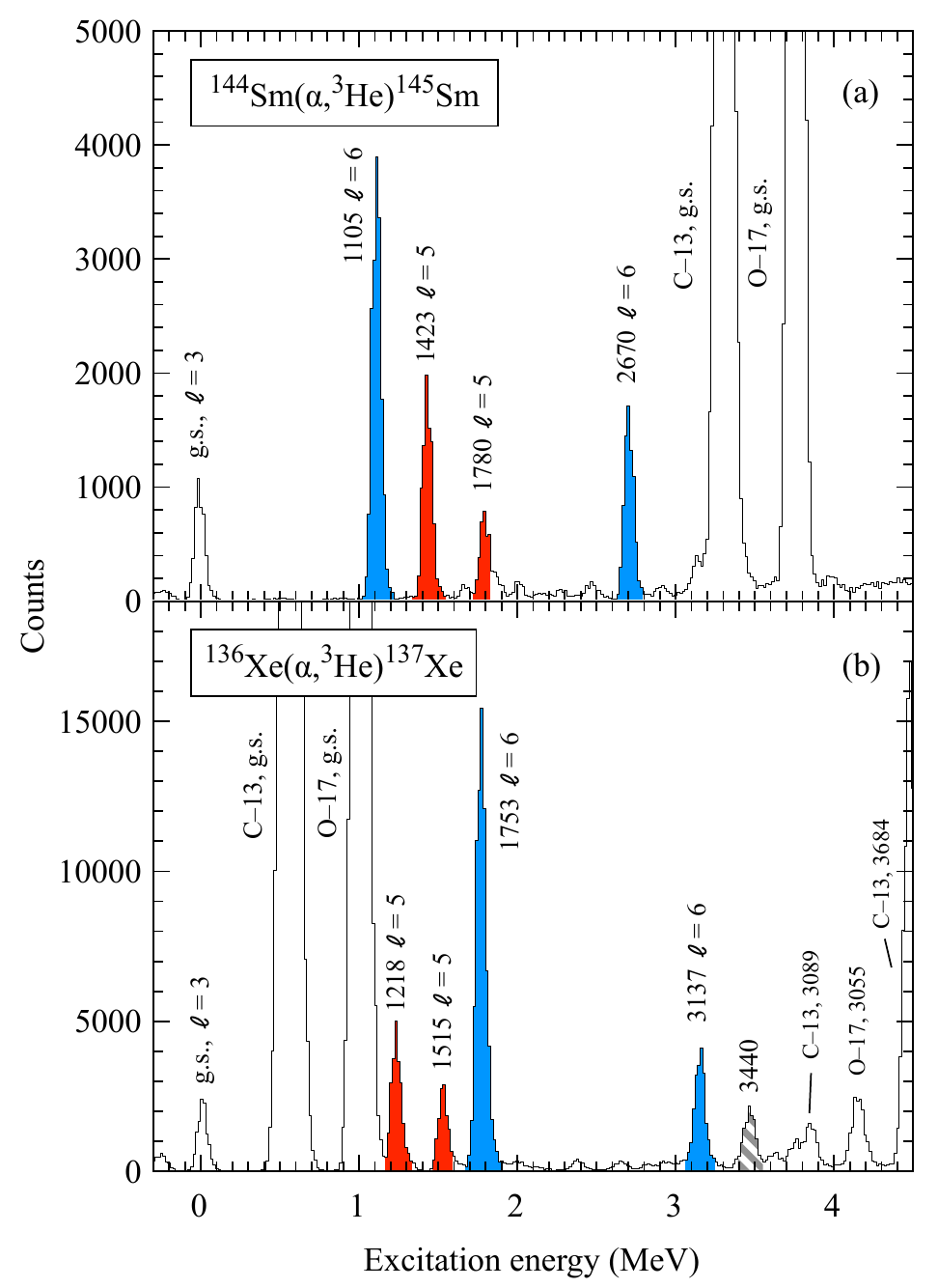}
\caption{\label{fig1} Outgoing $^3$He spectra following the ($\alpha$,$^3$He) reaction on isotopes of $^{144}$Sm (a) and $^{136}$Xe (b) from $E_{\alpha}=100$~MeV and $\theta_{\rm lab}=6^{\circ}$. The $9/2^-$ ($\ell=5$) states are shaded red and $13/2^+$ states ($\ell=6$) in blue. The 3440-keV state (hatched) is tentatively assigned as a third 13/2$^+$ state.}
\end{figure}

The spectra from the ($\alpha$,$^3$He) reaction on $^{144}$Sm and $^{136}$Xe are shown in Fig.~\ref{fig1} for a spectrometer angle $\theta_{\rm lab}=6^{\circ}$. The spectra are dominated by states populated via $\ell=5$ and $\ell=6$ transfer. Aside from these high-$j$ states and the ground state, which contains more-or-less all of the $\nu1f_{7/2}$ strength~\cite{Ipson73}, essentially no other states are populated with significant yield. The $Q$-value resolution was 65--70~keV full-width at half-maximum. Contaminants, resulting from reactions on carbon and oxygen, lie above the region of interest in $^{145}$Sm. For Xe, the contaminant peaks fall just below the high-$j$ states of interest at $\theta_{\rm lab}=4^{\circ}$ and $6^{\circ}$, and slowly move across this region as the angle becomes larger. Beyond $4^{\circ}$ and $6^{\circ}$, only at $18^{\circ}$ were all high-$j$ states unobscured.

The Sm spectrum is, as expected, similar to that of Fig.~1 of Ref.~\cite{Kay08} which shows the same reaction but carried out at a lower energy of $E_{\alpha}=51$~MeV. As mentioned in the introduction, this pattern is observed in other $N=83$ isotones. The corresponding states are clearly seen in $^{137}$Xe. The 3137-keV 13/2$^+_2$ state, observed recently for the first time in the study of Reviol {\it et al.}~\cite{Reviol16}, is clearly seen. 

\begin{figure}
\centering
\includegraphics[scale=0.85]{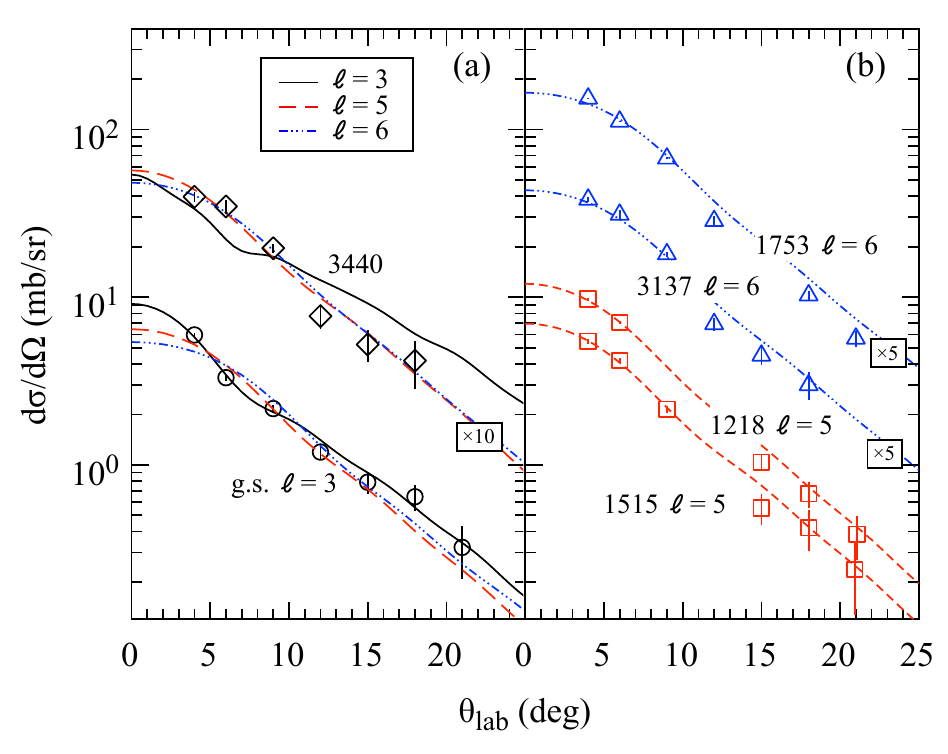}
\caption{\label{fig2} Angular distributions for states populated in the $^{136}$Xe($\alpha$,$^3$He)$^{137}$Xe reaction at 100~MeV. Panel (a) shows the ground-state $\ell=3$ transition (circles) and the 3440-keV state with $\ell=5$ or $\ell=6$ transfer ($\times$10, diamonds) and (b) the known 9/2$^-$ (squares) and 13/2$^+$ ($\times$5, triangles) states. The curves are DWBA calculations fitted to the data for the respective $\ell$ transfers.}
\end{figure}

The angular distributions for the states populated in $^{137}$Xe are shown in Fig.~\ref{fig2}. Although the choice of reaction and incident beam energy lends itself to exquisite selectivity for populating high-$\ell$ states, the angular distributions are relatively featureless, particularly for any differences in the $\ell=5$ and $\ell=6$ shapes. As such, making robust $\ell$-value assignments is not possible. We rely on the previous spin and parity assignments for the 1218-keV and 1515-keV, 9/2$^-$ states. Note, the latter was previously reported to lie at an excitation energy of 1590(20) keV~\cite{Kay11}. This was observed as part of an unresolved  doublet with a known $\ell=3$ state at 1534~keV via the ($d$,$p$) reaction in inverse kinematics, where the energy resolution, statistics, and momentum matching for $\ell=5$, were poor. Here, the state is cleanly populated and the excitation energy unambiguous. Due to the selectivity of heavy-ion transfer, the $\ell=5$, 9/2$^-_1$ strength was only weakly populated in previous works~\cite{Allmond12,Reviol16}, and the 9/2$^-_2$ excitation was not observed. The two 13/2$^+$ states, at 1753~keV and 3137~keV, have also been previously assigned. The experimental angular distributions for these $\ell$ transfers agree well with the shapes calculated with the finite-range distorted wave Born approximation (DWBA) using the code Ptolemy~\cite{ptolemy}. The ground-state $\ell=3$ is also well reproduced by the DWBA. While similar to the $\ell=5$ and $\ell=6$ shapes, the  $\chi^2$/dof of the fits for these shapes strongly favor $\ell=3$, being 1.05, 6.4, and 11.2 for $\ell=3$, 5, and 6, respectively.

An additional peak was observed at 3440~keV, which was populated with a cross section approximately seven times less than that of the lowest-lying 13/2$^+$ state. The angular distribution for the 3440-keV state is consistent with $\ell=5$ or $\ell=6$ transfer, but not with $\ell=1$ (which is poorly momentum matched in this reaction and would lead to unphysical values for the overlaps) or $\ell=3$. Previous ($d$,$p$) studies, such as those of Ref.~\cite{Kay11}, had neither sufficient statistics nor resolution to resolve states above approximately 3~MeV in excitation energy. As we will discuss, we tentatively assign the 3440-keV state as $j^{\pi}=13/2^+$. Two additional states with cross sections a factor of 10 times less than the dominant $\ell=6$ state were observed above the 3440-keV state. These may also have high spin. All cross sections are included in the Supplemental Material~\cite{supmat}.

Different parameterizations of optical-model potentials were explored for the DWBA calculations. The availability of global optical-model parameterizations for $\alpha$ particles in this energy regime is quite limited compared to lighter ions. We used those of Refs.~\cite{Nolte87,Su15} along with those derived from scattering experiments optimized for $\sim$100~MeV energies and on nuclei close in mass to the ones studied here. The most favorable being that of Ref.~\cite{Perry81} who derived parameters from the scattering of $\sim$80-MeV $\alpha$ particles on $^{208}$Pb. Other works demonstrated that these same parameters reproduce angular distributions for the ($\alpha$,$^3$He) reaction at 80 MeV~\cite{Gales85} and 100 MeV~\cite{Duffy86} on $^{144}$Sm.

\begin{table}
\caption{\label{tab1} Excitation energies (to the nearest keV unless otherwise stated), differential cross sections ($\sigma$), and normalized spectroscopic factors ($S$) for states populated via $\ell=5$ and $\ell=6$ transfer in the ($\alpha$,$^3$He) reaction on $^{136}$Xe at $\theta_{\rm lab}=6^{\circ}$. }
\newcommand\T{\rule{0pt}{3ex}}
\newcommand \B{\rule[-1.2ex]{0pt}{0pt}}
\begin{ruledtabular}
\begin{tabular}{ccccc}
 $E$ (keV)\B & $\ell$ & $j^{\pi}$ & $\sigma_{6^{\circ}}$ & $S$  \\
\hline
1218\T & 5 & 9/2$^-_1$  & 6.45(4) & 0.51(4) \\
\T  1515(5)\footnote{This energy of this state was previously estimated to be at 1590(20)~keV in Ref.~\cite{Kay11}.} & 5 & 9/2$^-_2$ & 3.67(3) & 0.29(3) \\
\T  1753 & 6 &13/2$^+_1$ & 21.0(1) & 0.78(7) \\
\T\B  3137(1) & 6 & 13/2$^+_2$ & 5.62(4) & 0.22(2)  \\
\T\B  3440(15) & (6) & (13/2$^+_3$)  & 2.97(3) & 0.12(2)  \\
\end{tabular}
\end{ruledtabular}
\end{table}

For the $^3$He parameterizations, numerous global optical-model potentials have been derived from data at these masses and incident energies. We explored those from Refs.~\cite{Trost87,Pang09,Xu11}. Overall, nine different combinations of parameters were used. For the bound-state wave function we used  parameterizations derived from the Green's function Monte Carlo calculations of Brida {\it et al.}~\cite{Brida11}. The target bound-state wave function was generated in the conventional way, using a Woods-Saxon potential with the depth varied to reproduce the binding energy of the transferred nucleon. The radial and diffuseness parameters of the potential were $r_0=1.28$~fm and $a=0.65$~fm and the spin-orbit potential was taken as $V_{\rm so}=6$~MeV, $r_{\rm so}=1.10$~fm, and $a_{\rm so}=0.65$~fm.

The spectroscopic factors quoted in Table~\ref{tab1} were extracted from the $\theta_{\rm lab}=6^{\circ}$ cross sections. The relative spectroscopic factors were derived from a common normalization \mbox{$N_j$, such that $N_j=\sum(2j+1)C^2S_j/(2j+1)$~\cite{Schiffer12}}. Across the nine different combinations of optical-model parameterizations, the $\theta_{\rm lab}=4^{\circ}$ and $\theta_{\rm lab}=6^{\circ}$ data, and the two different $j$ values, a normalization for the $^{137}$Xe states of 0.52(3) was found, where the quoted uncertainty is the rms spread. This spread, though small, also includes different summed strengths based on whether the previously unobserved 3440-keV state is assigned $j^{\pi}=9/2^-$ or $j^{\pi}=13/2^+$. The spectroscopic factors in Table~\ref{tab1} include the 13/2$^+_3$ state in the normalization.

The normalized spectroscopic factors for states populated in $^{145}$Sm agree well with those reported in Ref.~\cite{Kay08}. The resulting centroids for $^{145}$Sm derived from the current data are $\epsilon_{h_{9/2}}=1531(10)$~keV and $\epsilon_{i_{13/2}}=1589(10)$~keV (cf.\ Table~\ref{tab2}). These values are within 10~keV of those calculated in Ref.~\cite{Kay08}.  In the remaining discussion we adopt the spectroscopic factors, centroids, and mixing matrix elements given in Ref.~\cite{Kay08} for $^{145}$Sm as a matter of convenience, as they also include a consistent analysis of the same properties for $^{143}$Nd, $^{141}$Ce, and $^{139}$Ba.

\begin{figure}
\centering
\includegraphics[scale=0.85]{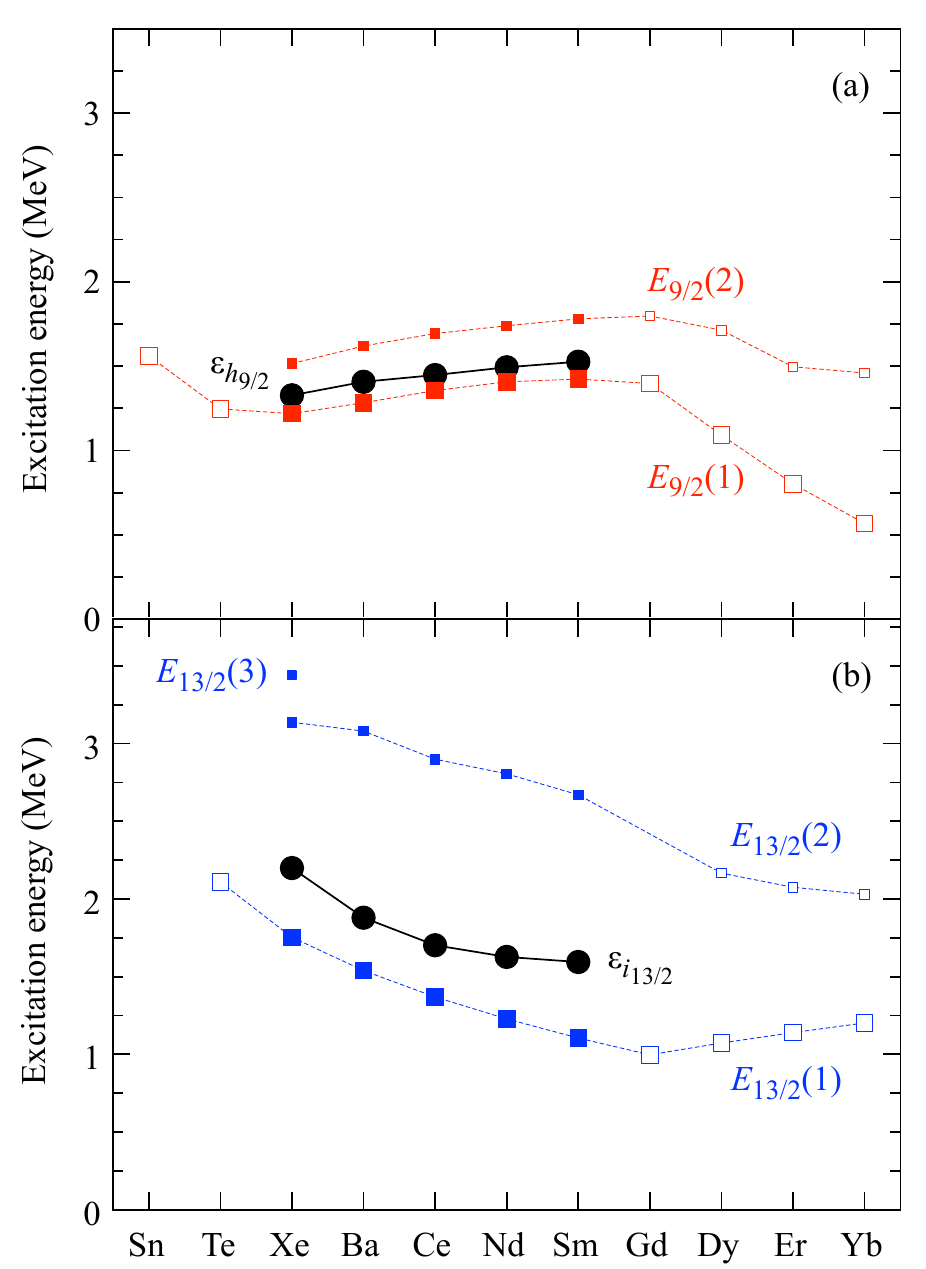}
\caption{\label{fig3} The excitation energies of the 9/2$^-$ (a) and 13/2$^+$ (b) states at $N=83$. The solid symbols are states for which spectroscopic factors from the ($\alpha$,$^3$He) reaction are available. A third tentatively assigned 13/2$^+$ state in $^{137}$Xe is also shown. The $0h_{9/2}$ and $0i_{13/2}$ centroids, $\epsilon_{h_{9/2}}$ and $\epsilon_{i_{13/2}}$, are also shown as black dots.}
\end{figure}

\begin{figure}
\centering
\includegraphics[scale=0.85]{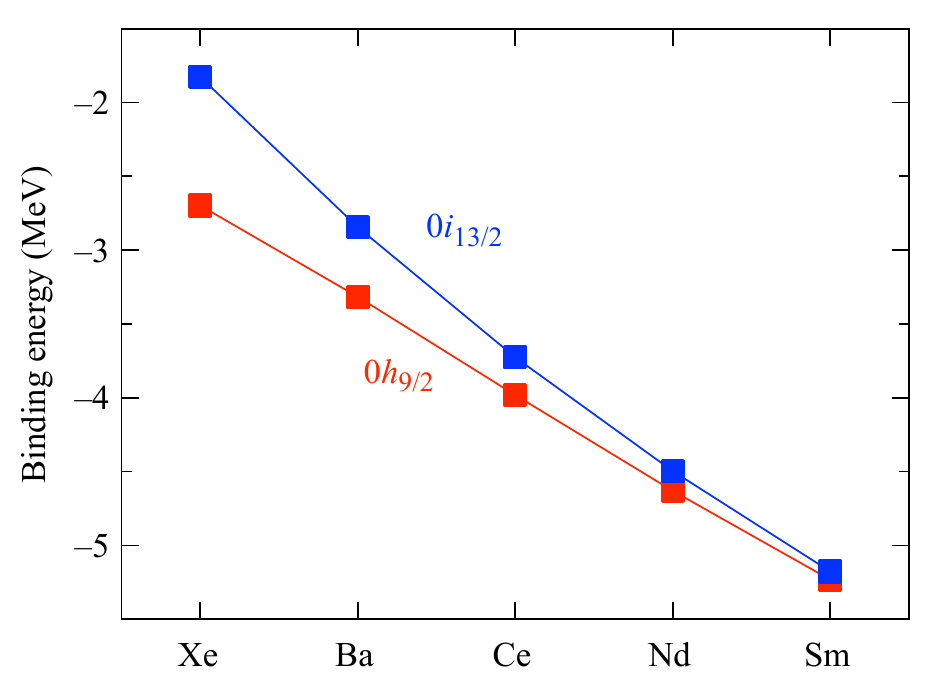}
\caption{\label{fig4} Binding energies of the neutron $0h_{9/2}$ and $0i_{13/2}$ excitations, corresponding to the centroids in Fig.~\ref{fig3}, as a function of proton number.}
\end{figure}

\begin{table}
\caption{\label{tab2} Centroids, in MeV, for the $0h_{9/2}$ and $0i_{13/2}$ excitations from $54<Z<62$. Those for $56<Z<62$ are from Ref.~\cite{Kay08}. The uncertainties are less than $\pm$0.02~MeV unless stated otherwise.}
\newcommand\T{\rule{0pt}{3ex}}
\newcommand \B{\rule[-1.2ex]{0pt}{0pt}}
\begin{ruledtabular}
\begin{tabular}{ccc}
Isotope \B & $\epsilon_{h_{9/2}}$ & $\epsilon_{i_{13/2}}$  \\
\hline
$^{137}$Xe\T  & 1.327(10)  & 2.206(20)\footnote{Includes the tentative 13/2$^+_3$ state.} \\
$^{139}$Ba\T  & 1.407(10) & 1.879(24)  \\
$^{141}$Ce\T  & 1.447(10) & 1.702 (52) \\
$^{143}$Nd\T  & 1.493(5) & 1.627(31) \\
$^{145}$Sm\T\B & 1.526(10) & 1.594(29) \\
\end{tabular}
\end{ruledtabular}
\end{table}

The relative spectroscopic factors from the $^{136}$Xe($d$,$p$) study carried out at 10~MeV/u in inverse kinematics~\cite{Kay11} agree within the quoted uncertainties with those of the present study. This bodes well for future ($d$,$p$) studies at comparable energies on isotopes of Te and Sn, where the $\alpha$-induced reactions are likely to prove challenging.

As mentioned above, the state at 3440~keV was unexpected and its relatively large cross section and angular distribution suggest a high-$\ell$ value. There is little to discriminate between $\ell=5$ and $\ell=6$ angular distributions at this incident beam energy (as seen in Fig.~\ref{fig2}). The systematics of the 9/2$^-$ and 13/2$^+$ excitations are shown in Fig.~\ref{fig3}. Were the 3440-keV state to have a $j^{\pi}=9/2^-$ assignment, it would have a spectroscopic factor $\sim$0.3 and the resulting centroid would lie at approximately 1885~keV, some 560~keV higher in excitation energy. This would be at odds with the  systematics shown in Fig.~\ref{fig3}. An assignment of $j^{\pi}=13/2^+$ is perhaps most likely, with a spectroscopic factor of 0.12(2) as quoted in Table~\ref{tab1}. With the 3440~keV state included in the normalization as $j^{\pi}=13/2^+$, and subsequent calculations of the centroid, it only shifts the energy of the $0i_{13/2}$ orbital by $\sim$140~keV. We searched for evidence of the 3440-keV state in other works, the most promising one being that of Reviol {\it et al.}~\cite{Reviol16}. This state could possibly decay via a 1687-keV $\gamma$-ray transition to the 13/2$^+_1$ state or to the 13/2$^+_2$ state via a 303-keV $\gamma$ ray. However, the statistics in these data are limited and there appears to be no clear evidence for such  decays~\cite{Reviol16,Reviol17}.

The uncertainties on the absolute cross sections are of the order of 30\% as discussed in Sec.~\ref{sec2}. This implies an uncertainty on the absolute spectroscopic factors of about the same order, around 30-40\%, which is dominated by the uncertainties in the cross sections. However, the relative spectroscopic factors have comparably smaller uncertainties, around 5\%, as seen in the rms spread of the normalizations using different optical-model parameterizations. The uncertainties shown in Table~\ref{tab1} for the cross sections are those of the statistical uncertainties and those related to the fitting procedure. Due to the high statistics of the recorded yields, these are of the order of 1\% for the relative cross sections. The normalized spectroscopic factors have an uncertainty of approximately 5--10\%, which reflects the variation between the different optical-model parameterizations.

\section{Discussion}

\subsection{Single-particle energies and the tensor force}

The $\nu0h_{9/2}$ and $\nu0i_{13/2}$ single-particle energy centroids at $^{137}$Xe are calculated to be $\epsilon_{h_{9/2}}=1327(10)$~keV and $\epsilon_{i_{13/2}}=2206(20)$~keV.  The latter includes the probable ``third 13/2$^+_3$ state'' at 3440~keV. We include this in the remainder of the discussion. The uncertainties, as presented in the figures and the tables, assume this state is correctly assigned, i.e., they do not include the $\sim$140~keV difference between the centroid were the assignment of the spin-parity for the 3440~keV state be in error. The centroids, $\epsilon_{h_{9/2}}$ and $\epsilon_{i_{13/2}}$, are reconstructed from the fragments as their centers of gravity, where $\epsilon_j=\sum E_i S_j/\sum S_j$. They are shown in Fig.~\ref{fig3} along with other data. The binding energies of these centroids are shown in Fig~\ref{fig4}. The centroids move closer together in binding energy by about 0.8~MeV going from Xe to Sm, with the binding of the  $\nu0i_{13/2}$ orbital increasing more rapidly with increasing proton number than that of the  $\nu0h_{9/2}$ orbital.

The $\pi0g_{7/2}$ and $\pi1d_{5/2}$ orbitals are filling above $Z=50$. The tensor interaction is attractive when the proton and neutron orbits have a different spin orientation with respect to their orbital angular momentum ($\ell+1/2$ and $\ell-1/2$, or vice versa) and repulsive when the spin orientation is the same. Its magnitude increases with increasing $\ell$ and is greatest between orbitals with zero nodes. In the present case, the $\pi0g_{7/2}$, $\nu0h_{9/2}$, and $\nu0i_{13/2}$ states are nodeless and have significant radial overlap.

\begin{figure}[h!]
\centering
\includegraphics[scale=0.85]{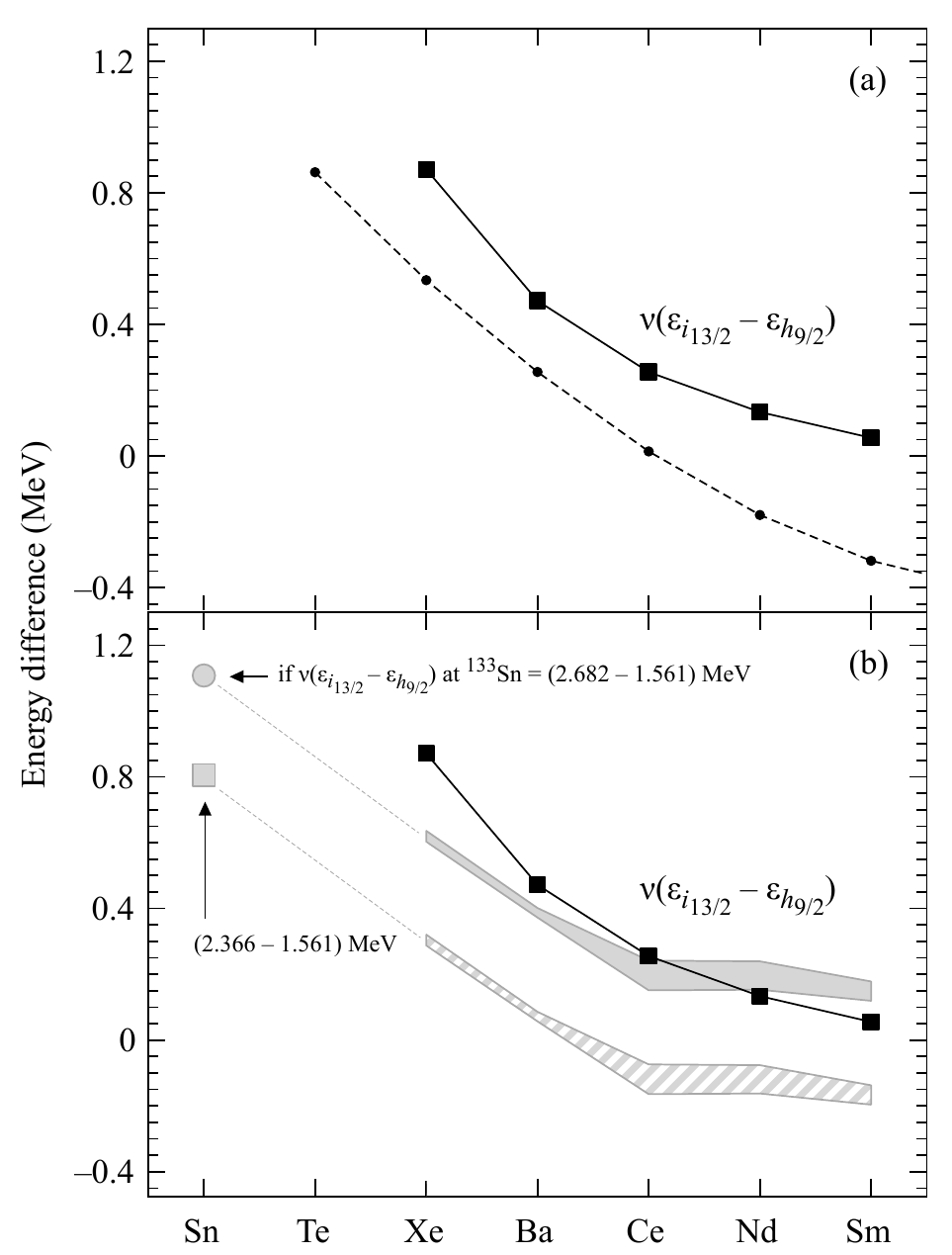}
\caption{\label{fig5} (a) The difference in the centroids of the $0i_{13/2}$ and $0h_{9/2}$ single-neutron strength (where data for Ba, Ce, Nd, and Sm are from Ref.~\cite{Kay08}) are shown as the black squares connected by a solid line. The dashed line is the energy difference between the 13/2$^+_1$ and 9/2$^-_1$ states for the $N=83$ isotones between Te and Sm~\cite{nndc}. In Panel (b) the black squares connected by the solid line are the experimental differences as in (a). The shaded and hatched areas represent calculations based on the tensor interaction~\cite{Otsuka05}, anchored to estimates of the $0i_{13/2}$ single-particle energy in $^{133}$Sn, based on two different predictions. The shaded region is anchored to the gray circle at $Z=50$, which is based on the average value of $\epsilon_{i_{13/2}}$ from Urban {\it et al.}~\cite{Urban99} and Korgul {\it et al.}~\cite{Korgul15}, 2.682~MeV. The location of the $\epsilon_{h_{9/2}}$ is taken as 1.561~MeV, the location of the observed 9/2$^-_1$ state~\cite{nndc}. The hatched area represents a similar calculation but anchored to the gray square at $Z=50$ which uses the estimate of $\epsilon_{i_{13/2}}=2.366$~MeV of \mbox{Reviol {\it et al.}~\cite{Reviol16}}. The width of the shaded region reflects the uncertainties in the proton occupancies~\cite{Wildenthal71,Entwisle16}. }
\end{figure}

The evolution of a given single-particle energy with respect to another can be expressed as $\epsilon_{j\nu}=\epsilon_{{\rm initial},j\nu}+\sum V_M(j_{\nu}j'_{\pi})N_{j\pi}$, where $\epsilon_{{\rm initial},j\nu}$ is a starting value.  $V_M(j_{\nu}j'_{\pi})$ is the magnitude of the monopole interaction from the tensor force, which has the effect of decreasing the separation between the $0i_{13/2}$ and $0h_{9/2}$ orbitals by 0.18~MeV per each proton added to the $\pi0g_{7/2}$ orbital ($N_{j\pi}$) and increasing the separation by 0.005~MeV per each proton added to the $\pi1d_{5/2}$ orbital~\cite{Kay08}. This effect is shown by the shaded and hatched areas in Fig.~\ref{fig5}. The proton occupancies, $N_{j\pi}$, are taken from Wildenthal {\it et al.}~\cite{Wildenthal71} and Entwisle {\it et al.}~\cite{Entwisle16}, who studied all the stable even $N=82$ isotones via proton transfer. The widths of the shaded and hatched regions reflect the uncertainties in these proton occupancies. The shaded and the hatched regions show the difference in the single-particle energies as described by the tensor interaction and known proton occupancies using two different predictions for the separation of the  $0i_{13/2}$ and $0h_{9/2}$ single-particle energies at $^{133}$Sn.

For both the shaded and the hatched region in Fig.~\ref{fig5}(b), the $0h_{9/2}$ energy in $^{133}$Sn is taken to be the observed 9/2$^-_1$ state at 1561~keV~\cite{nndc}. This is likely a fair assumption as the fragmentation is expected to be minimal with only one neutron outside the doubly magic core, and with the 2$^+$ excitation above 4~MeV in excitation energy, nearly 3~MeV higher than in $^{134}$Te. Most previous predictions for $\epsilon_{i_{13/2}}$ at $^{133}$Sn have placed it above the neutron separation energy of \mbox{$S_n=2402(4)$~keV}. The $0i_{13/2}$ energy for $^{133}$Sn that results in the shaded region in Fig.~\ref{fig5}(b) is $\epsilon_{i_{13/2}}=2682$~keV, the average of those given by Urban {\it et al}~\cite{Urban99} and by Korgul {\it et al.}~\cite{Korgul15}. With these values for the single-particle energies in $^{133}$Sn, the agreement between the calculated differences in the centroids as a function of proton number, as indicted by the shaded area, and the experimental centroids is remarkably good. Using a value of 2366~keV for the $0i_{13/2}$ energy~\cite{Reviol16} shifts the calculated differences down by $\sim$370~keV, as shown by the gray hatched area. From studies of heavy-ion transfer reactions, Allmond {\it et al.}~\cite{Allmond14} observed a 2792-keV $\gamma$-ray, for which it was tempting to infer this as a candidate for a 13/2$^+$ state being consistent with the Urban {\it et al.} within their quoted uncertainties. However, no assignment was made based on other considerations.

\subsection{Trends in mixing matrix elements}\label{mixing}

As noted in previous works (e.g.,~\cite{Booth74,Heyde75,Isacker79,Trache93,Oros95,Kay08}, among others), the $0h_{9/2}$ and $0i_{13/2}$ strength is split, apparently mixing with the weak-coupling state of the $\nu 1f_{7/2}$ coupled to the 2$^+_{\rm core}$ or 3$^-_{\rm core}$ states. 

\begin{figure}
\centering
\includegraphics[scale=0.85]{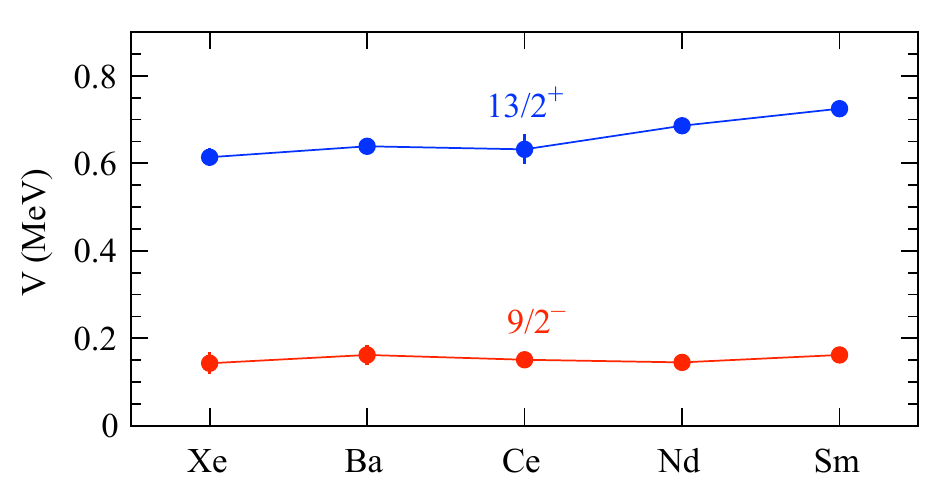}
\caption{\label{fig6} The mixing matrix elements. The error bars reflect the uncertainties in the derived values.}
\end{figure}

The mixing matrix elements between the single-particle and weak-coupling state can be extracted using a simple two-level mixing model, which we describe in the Appendix. The mixing matrix elements are shown in Table~\ref{tab3} and Fig.~\ref{fig6} for $54<Z<62$. In this example, the $j^{\pi}=9/2^-$ admixtures are assumed to be the $|0h_{9/2}\rangle$ and $|2^+,1f_{7/2}\rangle$, and  $j^{\pi}=13/2^+$ admixtures as $|0i_{13/2}\rangle$ and $|3^-,1f_{7/2}\rangle$, as in Ref.~\cite{Kay08}. The mixing matrix elements are relatively strong, especially for the $0i_{13/2}$ state. For the 9/2$^-$ state, the magnitude of the mixing matrix element is remarkably constant across these isotopes, with an average value of 0.153(9)~MeV, where the rms spread across the five isotopes is given in parentheses. For the 13/2$^+$ states it is a factor of $\sim$4 larger, with an average value of 0.66(5)~MeV. 

\begin{table}[h!]
\caption{\label{tab3} Mixing matrix elements ($V$) for $N=83$, \mbox{$54<Z<62$}, all in MeV.}
\newcommand\T{\rule{0pt}{3ex}}
\newcommand \B{\rule[-1.2ex]{0pt}{0pt}}
\begin{ruledtabular}
\begin{tabular}{ccc}
& \multicolumn{2}{c}{$V$} \\
\cline{2-3} 
 &  \T\B 9/2$^-$, $2^+_{\rm core}\otimes \nu f_{7/2}$ & 13/2$^+$, $3^-_{\rm core}\otimes \nu f_{7/2}$ \\
\hline
$^{137}$Xe\T  & 0.143(5) & 0.614(20)\footnote{Includes the tentative 13/2$^+_3$ state in the centroid of $\epsilon_{i_{13/2}}$.} \\
$^{139}$Ba\T  & 0.162(6) & 0.639(17) \\
$^{141}$Ce\T  & 0.151(5) & 0.632(34) \\
$^{143}$Nd\T  & 0.145(2) & 0.686(16) \\
$^{145}$Sm\T\B & 0.162(4) & 0.725(9) \\
\end{tabular}
\end{ruledtabular}
\end{table}

Another $j^{\pi}=13/2^+$ state at approximately the same excitation energy can also arise from the 13/2$^+$ member of the weak-coupling multiplet corresponding to the core 2$^+$ state and $\nu0i_{13/2}$ state, $|2^+,0i_{13/2}\rangle$, as was discussed in other works, e.g., Heyde {\it et al.}~\cite{Heyde75}. This lends further support to the assignment of the 3440-keV as 13/2$^+_3$. We note that Heyde {\it et al.}'s calculations of the excitation energy and amplitude of the three 13/2$^+$ states in $^{137}$Xe are remarkably close to what we observe in this measurement, although the same calculated properties for the heavier isotones are in poorer agreement with the experimental data. A third 13/2$^+$ state, most likely arising from $|2^+,0i_{13/2}\rangle$, was not seen in the heavier $N=83$ isotones in the work of Ref.~\cite{Kay08}. In that work the limit on observing such a state would be at the level of $\sim$5\% of the lowest-lying state. A description of the possible three 13/2$^+$ states would require a more complex three-level mixing calculation. 

\begin{figure}
\centering
\includegraphics[scale=0.85]{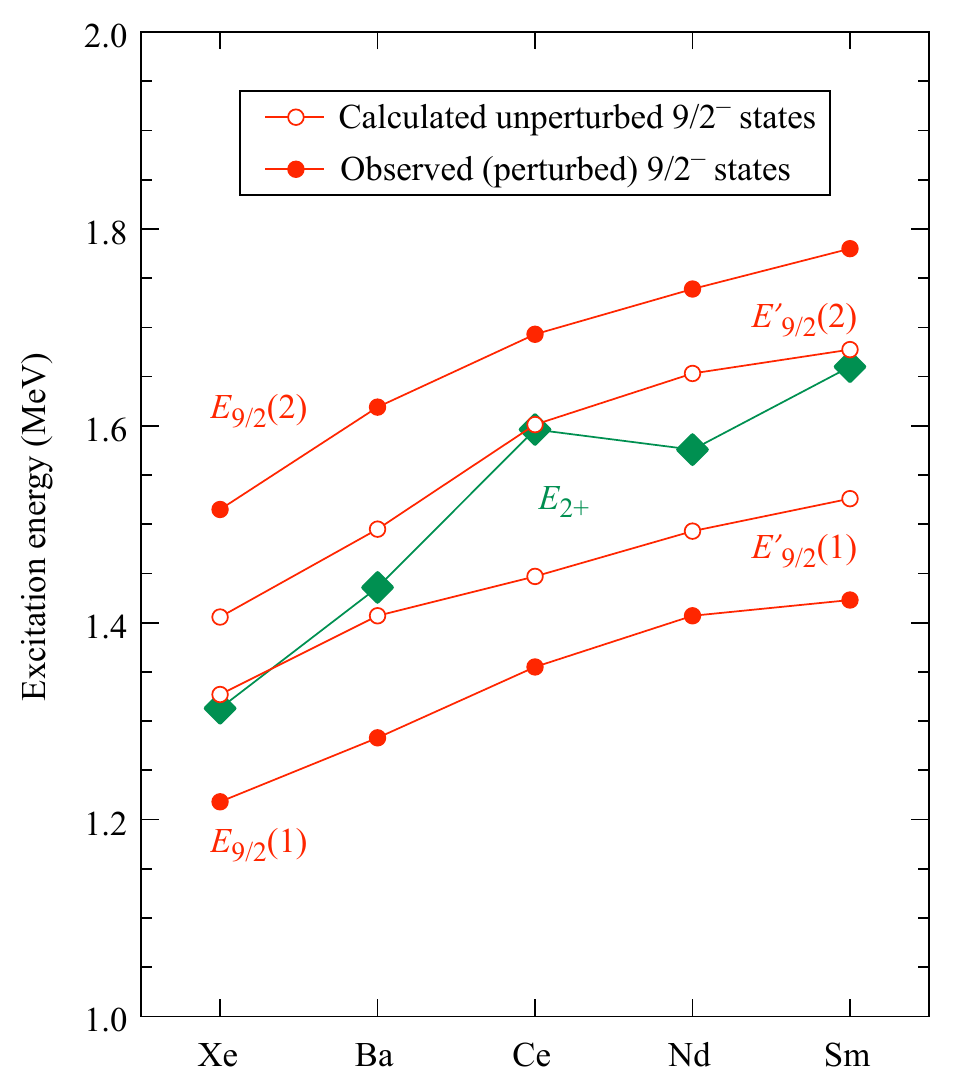}
\caption{\label{fig7} The excitation energy of the core 2$^+$ state and the perturbed ($E$) and unperturbed ($E'$) 9/2$^-$ excitations; $E'_{9/2}(1)\equiv\epsilon_{h_{9/2}}$. }
\end{figure}

\begin{figure*}
\centering
\includegraphics[scale=0.85]{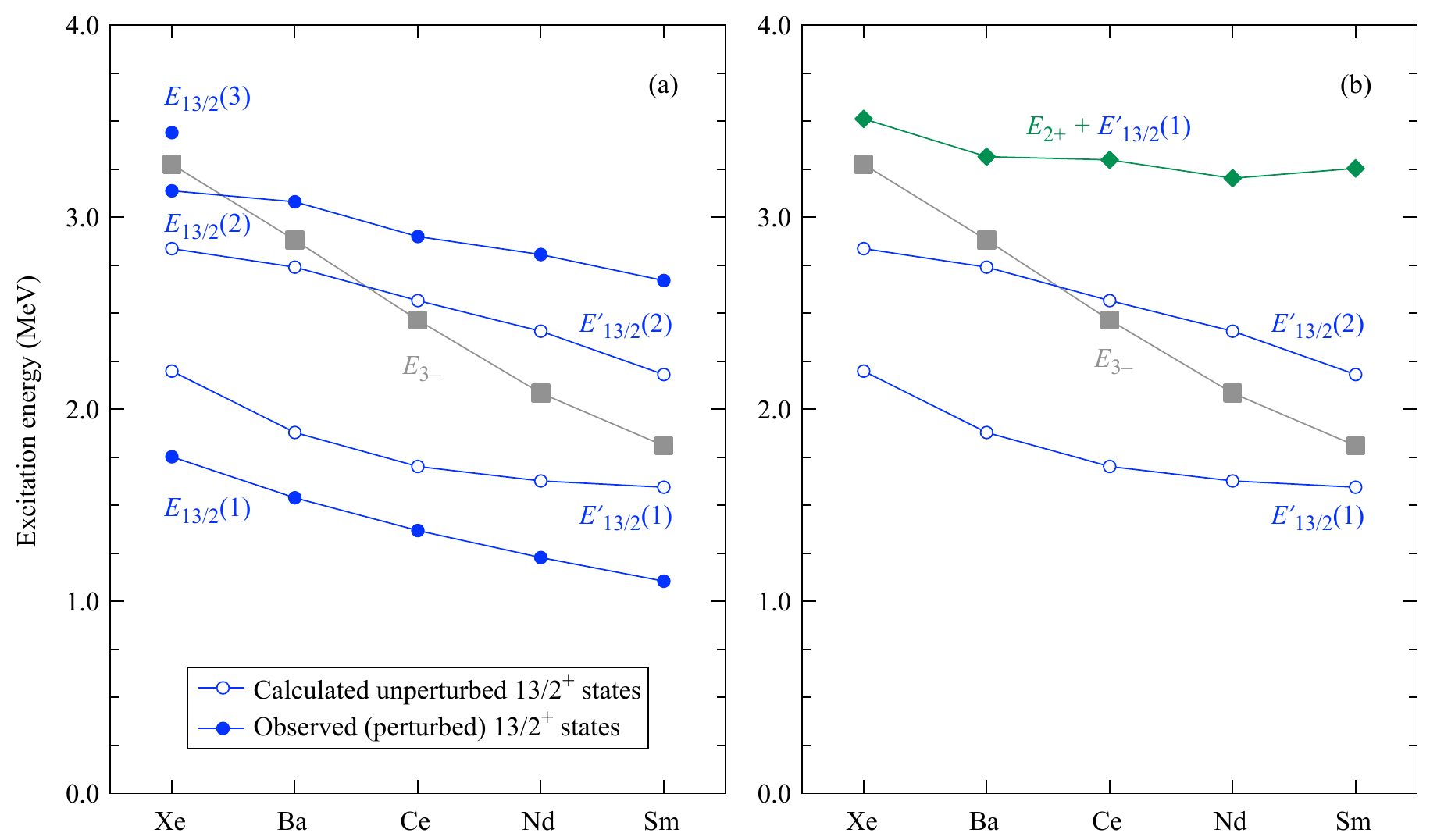}
\caption{\label{fig8} (a) The measured excitation energy ($E$) of the 13/2$^+$ states in $N=83$ nuclei from $54<Z<62$, the calculated unperturbed energies ($E'$), and the core ($N=82$) 3$^-$ energy. Panel (b) shows information on the core 2$^+$ states. Note, $E'_{13/2}(1)\equiv\epsilon_{i_{13/2}}$.}
\end{figure*}

In solving for the mixing matrix elements one also obtains the original unperturbed values of the ``single-particle state'' (the `centroid', that was already discussed) and the ``weak-coupling state''. Figures~\ref{fig7} and~\ref{fig8}(a) show these for the 9/2$^-$ and 13/2$^+$ states, respectively.

The unperturbed 9/2$^-$ energy tracks the 2$^+_{\rm core}$ very closely, being consistently within 0.2~MeV of it from $54<Z<62$. The energy of the 2$^+_{\rm core}$ state only changes by a few tenths of an MeV across that range. The relationship between the 3$^-_{\rm core}$ and the unperturbed 13/2$^+$ excitations is more startling. Over the same range the 3$^-_{\rm core}$ excitation energy changes by over an MeV. The excitation energy of the unperturbed 13/2$^+_2$ state is {\it below} the 3$^-_{\rm core}$ by $\sim$0.4~MeV at Xe, and at Sm it is almost 0.4~MeV {\it above} it.

Figure~\ref{fig8}(b) also shows the 2$^+_{\rm core}$ in relation to the 3$^-_{\rm core}$ and unperturbed 13/2$^+$ excitations. As $Z$ decreases from Sm to Xe, the 3$^-_{\rm core}$ state rises to an energy close to the sum of the 2$^+_{\rm core}$ state and the $\nu0i_{13/2}$ energies, again lending support to the notion that the 3440~keV  has \mbox{$j^{\pi}=13/2^+$}.

We note that the behavior of the $3^-$ excitations at $N=82$ appears to be unique across the chart of nuclides for semi-magic nuclei. For example, at $Z=20$, 50, and 82, and at $N=28$ and 50,  $3^-$ excitations are relatively constant in energy while crossing the doubly magic nuclei, unlike the 2$^+$ states. $^{132}$Sn is an exception, in that at $N=82$, the 3$^-$ excitation rises sharply from $\sim$2.6~MeV at $^{130}$Sn to 4.35~MeV at $^{132}$Sn, as one tracks it along $Z=50$. It then decreases monotonically from $^{132}$Sn to $^{146}$Gd, from 4.35 to 1.58~MeV, as one tracks it along $N=82$. Over this region, protons are filling the $\pi0g_{7/2}$ orbital and this trend may perhaps be attributed to the increasing contribution of protons from these states being excited to the $\pi0h_{11/2}$ orbits. The change in sign occurs approximately where these orbits are half filled. To our knowledge, no systematic shell-model studies of this excitation have been carried out over this region. Perhaps the 3$^-$ excitation is not as collective as in other regions?

\section{Conclusion}

The high-$j$, 9/2$^-$ and 13/2$^+$ single-neutron excitations outside of $^{136}$Xe have been studied via the ($\alpha$,$^3$He) reaction at an incident energy of 100~MeV. In this work, two prominent states with $j^{\pi}=9/2^-$ and two with $j^{\pi}=13/2^+$ were populated. This is in line with what has been observed in the heavier $N=83$ systems and can be interpreted as the result of weak coupling to core vibrational states. An additional weak fragment was observed, with a probable assignment of $j^{\pi}=13/2^+$.  The $\nu0h_{9/2}$ and $\nu0i_{13/2}$ single-particle energies were determined to lie at 1327(10)~keV and 2206(20)~keV, respectively. They reflect a smooth continuation of the trends seen in the heavier $N=83$ isotones and are well described by the action of the tensor interaction based on the assumption of a $\epsilon_{i_{13/2}}\approx2680$~keV at $^{133}$Sn. While the ($d$,$p$) reaction is not the ideal probe of such states, at 10~MeV/u it can provide valuable information as evidenced by the results obtained in a recent inverse-kinematics study, which yielded comparable information in terms of accuracy. Such experiments with radioactive ion beams of $^{134}$Te and $^{132}$Sn at energies around 10~MeV/u are likely to be possible soon.

\section{Acknowledgments}

This measurement (E453) was performed at RCNP at Osaka University. The authors wish to thank the RCNP operating staff, and the outside participants wish to thank the local staff and administration for their hospitality and assistance. We acknowledge insightful discussions with Kris Heyde and Walter Reviol. We are indebted to John Greene of Argonne National Laboratory for preparing Sm targets for this experiment. This material is based upon work supported by the U.S. Department of Energy, Office of Science, Office of Nuclear Physics, under Contract Number DE-AC02-06CH11357, the Australian Research Council Discovery Project 120104176, and the UK Science and Technology Facilities Council. 

\appendix*
\section{}
\label{appendix}

\begin{figure}
\centering
\includegraphics[scale=0.5]{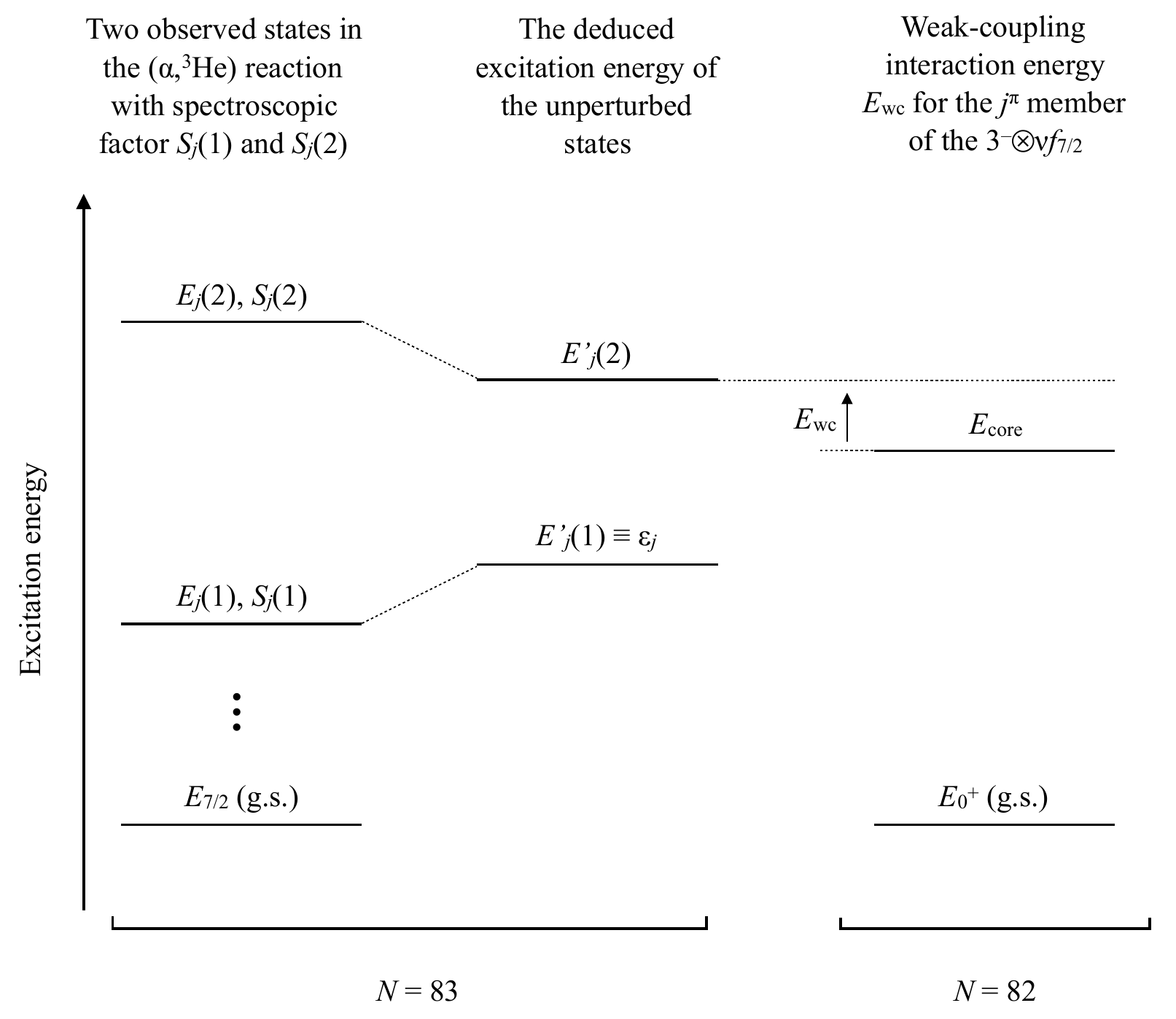}
\caption{\label{fig9} A schematic of the two-level mixing model used in the analysis in Sec.~\ref{mixing}. This is shown for 13/2$^+$ states---the same definitions apply for the 9/2$^-$ states. The energies, spectroscopic factors, and weak-coupling interaction energy are defined in the accompanying text.}
\end{figure}

A simple two-level mixing model was used to deduce the mixing matrix elements discussed in Sec.~\ref{mixing}. We describe the method for extracting these here, with notation that is consistent with the text and figures above. Figure~\ref{fig9} is a schematic showing the excitation energies of the perturbed (observed) states, $E_j$, and the unperturbed (deduced) states, $E'_j$ at $N=83$ and the core excitation at $N=82$. The relevant information for $N=83$ excitation energies, perturbed and unperturbed, and spectroscopic factors are given in Table~\ref{tab4} for $54\leq Z\leq62$.

The two wave functions for the observed states can be written as \mbox{$\Psi_1=\alpha\psi_1+\beta\psi_2$} and \mbox{$\Psi_2=-\beta\psi_1+\alpha\psi_2$}, where $\Psi$ are the actual wave functions and $\psi$ are the unperturbed wave functions. The amplitudes relate to the spectroscopic factors, where $S_1=\alpha^2$ and $S_2=\beta^2$. These are normalized such that $\alpha^2+\beta^2=1$. 

Using the notation of Fig.~\ref{fig9}, it follows that $E'_{j}(1)$, the unperturbed energy of the single-particle state, is given by
\begin{equation}
E'_{j}(1)=S_{j}(1)E_{j}(1)+S_{j}(2)E_{j}(2),
\end{equation}
and that of the weak-coupling state by
\begin{equation}
E'_{j}(2)=S_{j}(2)E_{j}(1)+S_{j}(1)E_{j}(2),
\end{equation}
where $E_j$ are the observed energies and $S_j$ are the normalized spectroscopic factors. The mixing matrix elements, $V$, are then calculated by
\begin{equation}
V=\frac{1}{2}\sqrt{\Delta E^2_{j}-\Delta E'^2_{j}},
\end{equation}
where $\Delta E^2=E_{j}(2)-E_{j}(1)$ and $\Delta E'^2=E'_{j}(2)-E'_{j}(1)$. The factor of $\frac{1}{2}$ is a matter of convention, sometimes it is expressed without this factor. $E_{\rm wc}$ in Fig.~\ref{fig9} is the weak-coupling interaction energy for the member of the multiplet that has the angular momentum $j$.

\begingroup
\squeezetable
\begin{table*}
\caption{\label{tab4} Energies to the nearest keV for excitations in the $N=83$ isotones between $54\leq Z\leq62$ that are relevant for the determination of the mixing matrix elements. Uncertainties are not given but can be found elsewhere in this work.}
\newcommand\T{\rule{0pt}{3ex}}
\newcommand \B{\rule[-2.3ex]{0pt}{0pt}}
\begin{ruledtabular}
\begin{tabular}{cccccccccccccccc}
& \multicolumn{7}{c}{$9/2^-,2^+_{\rm core}\otimes \nu f_{7/2}$ \T\B} & \multicolumn{7}{c}{$13/2^+,3^-_{\rm core}\otimes \nu f_{7/2}$ }  \\
\cline{2-8} \cline{9-15}
$^AX$\B\T  & $E_{9/2}(1)$ & $S_{9/2}(1)$ & $E_{9/2}(2)$ & $S_{9/2}(2)$ & $E'_{9/2}(1)$ & $E'_{9/2}(2)$ & $V$ & $E_{13/2}(1)$ & $S_{13/2}(1)$ & $E_{13/2}(2)$ & $S_{13/2}(2)$ & $E'_{13/2}(1)$ & $E'_{13/2}(2)$  & $V$ \\
\hline
$^{136}$Xe\T & 1218 & 0.43 & 1515 & 0.24 & 1327 & 1406 & 143 & 1753 & 0.84 & 3137 & 0.15 & 2206 & 2836 &  614\footnote{Like Table~\ref{tab3}, includes the tentative 13/2$^+_3$ state in the centroid of $\epsilon_{i_{13/2}}$.} \\
$^{138}$Ba\T  & 1283 & 0.70 & 1619 & 0.41 & 1407 & 1495 & 162 & 1539 & 0.60 & 3080 & 0.17 & 1879 & 2740 &  639 \\
$^{140}$Ce\T  & 1355 & 0.67 & 1693 & 0.25 & 1447 & 1601 & 151 & 1369 & 0.79 & 2899 & 0.22 & 1702 & 2566 &  632 \\
$^{142}$Nd\T  & 1407 & 0.83 & 1739 & 0.29 & 1493 & 1653 & 145 & 1228 & 0.65 & 2805 & 0.22 & 1627 & 2406 &  686 \\
$^{144}$Sm\T  & 1423 & 0.84 & 1780 &  0.34 & 1526 & 1677 & 162 & 1105 & 0.66 & 2670 & 0.30 & 1594 & 2181 & 725 \\
\end{tabular}
\end{ruledtabular}
\end{table*}
\endgroup


\end{document}